\documentclass[journal]{IEEEtran_kit}

\usepackage{epstopdf}
\usepackage{cite}
\usepackage{amsmath}
\usepackage{amsfonts}
\usepackage{mathtools}
\usepackage{algorithmic}
\usepackage{bbm}
\usepackage{caption}
\usepackage{array}
\usepackage{booktabs}
\usepackage{diagbox}
\usepackage{multirow}
\usepackage{multicol}
\usepackage[subfigure]{graphfig}
\usepackage{float}
\usepackage{array}
\usepackage[utf8]{inputenc}
\usepackage{color}
\DeclareCaptionFont{blue}{\color{blue}}

\begin{document}
\title{Deep Learning for Joint Channel Estimation and Signal Detection in OFDM Systems}
\author{\authorblockN{Xuemei Yi and Caijun Zhong}
\thanks{X. Yi, and C. Zhong are with the College of Information Science and Electronic Engineering, Zhejiang University, Hangzhou, China, and also with the Zhejiang Provincial Key Laboratory of Information Processing, Communication and Networking, Hangzhou. (email: caijunzhong@zju.edu.cn).}
}
\maketitle

\begin{abstract}
In this paper, we propose a novel deep learning based approach for joint channel estimation and signal detection in orthogonal frequency division multiplexing (OFDM) systems by exploring the time and frequency correlation of wireless fading channels. Specifically, a Channel Estimation Network (CENet) is designed to replace the conventional interpolation procedure in pilot-aided estimation scheme. Then, based on the outcome of the CENet, a Channel Conditioned Recovery Network (CCRNet) is designed to recover the transmit signal. Experimental results demonstrate that CENet and CCRNet achieve superior performance compared with conventional estimation and detection methods. In addition, both networks are shown to be robust to the variation of parameter chances, which makes them appealing for practical implementation.
\end{abstract}


\IEEEpeerreviewmaketitle
\section{Introduction}
Orthogonal frequency division multiplexing (OFDM) is a widely used modulation scheme in modern wireless systems. The performance of OFDM systems depends heavily on the adopted channel estimation and signal detection methods. As such, substantial research efforts have been devoted to the design of efficient channel estimation and reliable signal detection approaches, see \cite{hsieh1998channel,mahmoud2008kalman} and references therein.

In addition to the conventional model based channel estimation and signal detection methods appeared in the past few decades, a new artificial intelligence (AI) based paradigm, which employs deep neural networks (DNN) to perform channel estimation and signal detection, has recently emerged as a promising alternative. In \cite{H.Ye0}, a novel fully connected DNN (FC-DNN) is used to detect the data by treating the task of joint channel estimation and signal detection as a black box. Later on, over the air experiments were conducted to demonstrate the performance of the FC-DNN method in real environments \cite{P.Jiang}. More recently, meta learning was proposed for the channel estimation task in \cite{H.Mao}. Furthermore, AI-based design for CP-free OFDM system and OFDM with index modulation were studied in \cite{J.Zhang,T.Luong}, and novel deep learning architectures and design methodologies were proposed in for OFDM receiver under the constraint of one-bit complex quantization \cite{Jeff}.

Different from the aforementioned works, we propose a DL-based method for joint channel estimation and signal detection in OFDM systems. Specifically, a channel estimation network (CENet) is designed to replace the conventional interpolation mechanism by capitalizing on the image super resolution technique \cite{dai2019second}. Then, motivated by the conditioned image synthesis technology \cite{mao2019bilinear}, a novel Channel Conditional Recovery Network (CCRNet) is designed, which uses the output of CENet to recover the transmitted signal. Simulation results show that the proposed CENet can significantly improve the channel estimation accuracy, and the joint CENet and CCRNet yields better bit error rate performance compared to the conventional zero-forcing (ZF) detector and regularized zero-forcing (RZF) detector. Moreover, the proposed CENet and CCRNet exhibits good generalization ability and is quite robust to the variation of channel parameters such as operating signal to noise ratios (SNR).

\section{System Model}\label{model section}
\subsection{DL-aided OFDM Systems}
\begin{figure*}[thp]
	\centering
	\setlength{\belowcaptionskip}{-0.4cm}
	\includegraphics[height=7cm,width=14cm]{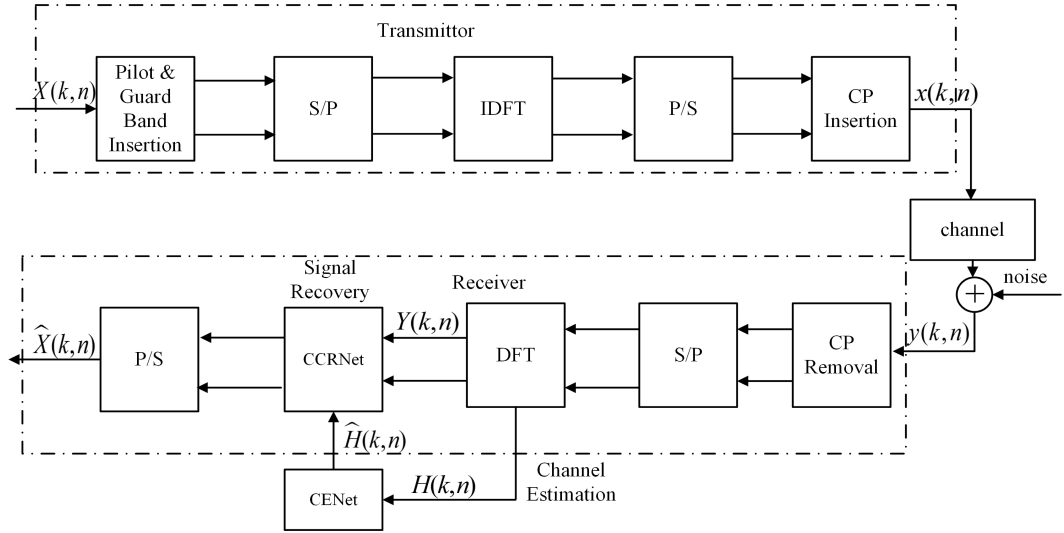}
	\caption{The DL-aided OFDM system architecture}
	\label{fig:system structure}
\end{figure*}
 We consider a single antenna OFDM system, and the proposed system architecture with deep learning based channel estimation and signal detection is illustrated in Fig. \ref{fig:system structure}. Without loss of generality, we consider an OFDM frame with $N$ time slots and $K$ subcarriers. Therefore, the received signal after removing the cyclic prefix (CP) and performing DFT at the subcarrier $k$ and the time slot $n$, $Y(k,n)$, can be expressed as:
\begin{equation}
Y(k,n) = H(k,n){\mkern 1mu} X(k,n)\; + {\mkern 1mu} W(k,n),
\end{equation}
where $k = 0,1, \ldots ,K$ and $n = 0,1, \ldots ,N$ are the subcarrier index and time index respectively. Also, $X(k,n) \in \mathbb{C}$ is the transmitted frequency symbol at subcarrier $k$ and time slot $n$, while $W(k,n) \in \mathbb{C}$ is the additive white Gaussian noise at subcarrier $k$ and time slot $n$, with zero mean and variance ${\sigma ^2}$, and $H(k,n)$ represents the frequency response at subcarrier $k$ and time slot $n$.

Assuming uniform power allocation over the subcarriers, the SNR of the system can be expressed as:
\begin{equation}
SNR = {\kern 1pt} {\kern 1pt} \frac{{{P_s}}}{{KF}} \bullet {\kern 1pt} \frac{1}{{{N_0}}},
\end{equation}
where ${P_s}$ is the total transmit power, ${{N_0}}$ denotes the noise power spectral density and $F$ is the subcarrier spacing.

\subsection{Conventional channel estimation }
In order to estimate the channel responses, pilot symbols are usually inserted into the frame with certain patterns. Once the channel responses at the pilot positions are obtained, some interpolation mechanism is used to get the entire channel response matrix. Denote ${\Omega _p}$ as the collection of two-dimensional (2D) pilot position indexes whereas the
number of pilot symbols is given by its cardinality $\left| {{\Omega _p}} \right| = {\kern 1pt} {N_p}$, and the channel coefficients ${H_P} \in {\mathbb C^{{N_P} \times 1}}$  at the pilot position can be expressed as:
\begin{equation}
{H_p}\; = \;\{ {H}({k_p},{n_p}),\forall ({k_p},{n_p})\; \in {\mkern 1mu} {\Omega _p}\},
\end{equation}
where ${n_p}$ and ${k_p}$ is the pilot index of subcarriers and time slots in an OFDM frame, respectively.

Then, using the well-established least square (LS) method or minimum mean squared error (MMSE), the estimated channel coefficients at the pilot positions, ${\hat H_p}$, can be expressed as:
\begin{equation}
\hat H_p^{LS} = {X_p}^{ - 1}{Y_p},
\end{equation}
or
\begin{equation}
\hat H_p^{MMSE} = \,{R_H}{[{R_H} + {({X_p}X_p^H)^{ - 1}}]^{ - 1}}\hat H_p^{LS},
\end{equation}
where ${Y_p} \in {^{{N_p} \times 1}}$ denotes the received pilot signal, ${X_p} \in {\mathbb C^{{N_p} \times {N_p}}}$  is a matrix which contains the known pilot symbols on its diagonal, and ${R_H}$ denotes the channel correlation matrix at the pilot-symbols with ${R_H} = E[{H_p}{H_p}^\dag ]$.

To this end, the entire channel coefficient matrix ${\hat H}$ can be obtained via interpolation schemes, such as linear interpolation or Gaussian interpolation \cite{hsieh1998channel}.

\subsection{Conventional Signal Detection}
With the estimated channel ${\hat H}$, zero-forcing detection \cite{Chen2004Zero} or
regularized zero forcing (RZF) \cite{alkamali2012regularized} can be applied to recover the transmit symbols. Hence, the post-processing signal at the subcarrier $k$ and the time slot $n$ can be expressed as
\begin{equation}
\hat X(k,n) = \frac{{Y\left( {k,n} \right)}}{{\hat H(k,n)}}
\end{equation}
or
\begin{equation}
\hat {X\,}{\rm{ = }}\,\left( {{{\hat H}^H}\hat H{\rm{ + }}\tau I} \right)\,{\widehat H^H}Y,
\end{equation}
where $\tau $ is the regularisation parameter.

\begin{figure}[tpb]
	\setlength{\abovecaptionskip}{0.cm}
	\setlength{\belowcaptionskip}{-0.4cm}
	\centering
	\includegraphics[height=3.5cm,width=7.8cm]{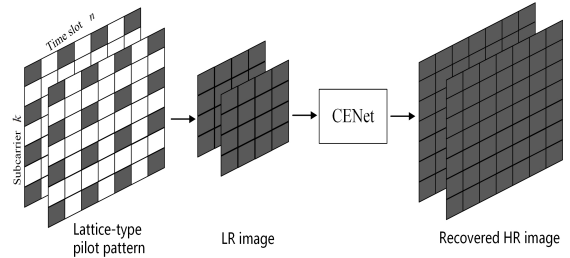}
	\caption{The proposed flow diagram for DL-based channel estimation.}
	\label{fig:CENet}
\end{figure}
\section{ DL-based channel estimation and signal detection}\label{channel estimation and signal detction}
\subsection{DL-based channel estimation}
As mentioned above, the entire channel coefficient matrix is computed via certain interpolation scheme, which nevertheless is blind to the underlying correlation structure of the channels, thereby leading to unsatisfactory performance\cite{mahmoud2008kalman}\cite{charrada2017analyzing}. Motivated by the recent advance of deep learning method, which has demonstrated superior ability of functional fitting, we propose to adopt a deep neural network to replace the interpolation procedure.

To exploit the sophisticated deep neural networks designed in the image processing community, we model the time-frequency channel matrix as a 2D image. Hence, the problem is to recover the entire channel matrix, which we denote as high-resolution (HR) image,  from small percentage of known channel coefficients at the pilot positions, which we denote as low-resolution (LR) image. Mathematically, the HR image $\hat{H}$ is obtained by:
\begin{equation}
\hat H = {G_{\sf CENet}}\left( {{\hat H_p^{LS}};\Theta } \right)
\end{equation}
where ${\hat H_p^{LS}}$ represents LR image obtained by the LS estimator, ${G_{\sf CENet}}$ denotes the CENet with $\Theta$ being the network parameters.

In this paper, we adopt the second-order attention network (SAN) architecture proposed in \cite{dai2019second}, since it has been shown to achieve state-of-the-art performance in super-resolution image recovery. Due to space limitation, the network structure and parameters of the SAN are omitted, only customized changes are highlighted here. Specifically, the size of the input and output channels are set to be $2$ since the channel response is a complex number (one channel for the real part and the other channel for the imaginary part). Meanwhile, the number of local-source residual attention group (LSRAG) modules is reduced to $15$ in order to improve the computational efficiency and avoid over-fitting.

Also, $L1$ norm is used as loss function to train the CENet, i.e.,
\begin{equation}
L(\Theta ){\mkern 1mu}  = {\mkern 1mu} \frac{1}{B}\sum\limits_{i = 1}^B {{{\left\| {{{\hat H}^i}{\mkern 1mu}  - {\mkern 1mu} {H^i}} \right\|}_1}},
\end{equation}
where $B$ is the batch size of a mini-batch training samples, $\left\{ {\widehat H^i} \right\}$ and $\left\{ {H^i} \right\}$ are the estimated channel responses and the corresponding ground-truth channel responses, respectively.

\subsection{DL-based signal detection}\label{signal detection}

\begin{figure}[t!]
	\centering
	\setlength{\belowcaptionskip}{-0.4cm}
	\includegraphics[height=8cm,width=8cm]{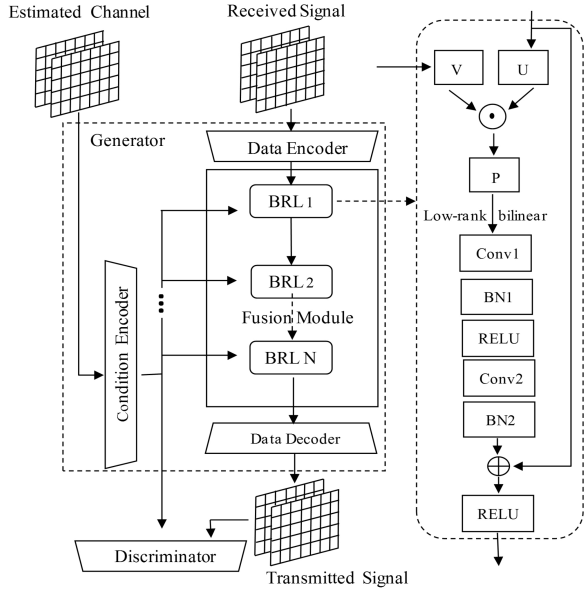}
	\caption{Overview of the proposed network architecture for signal recovery. Detail of the
		bilinear residual layer is presented in the dashed box.}
	\label{fig:CCRNet}
\end{figure}

In this section, we propose a DL-based signal detection
scheme. Since signal detection can be regarded as recovering the transmit signal from the received signal conditioned on the estimated channels, we propose to exploit the conditioned generative model for signal detection. Specifically, capitalizing on the conditioned image synthesis technique proposed in \cite{mao2019bilinear}, we design a channel condition recovery network (CCRNet) as illustrated in Fig.\ref{fig:CCRNet}. Hence, the recovered symbol $\hat{X}$ is given by
\begin{equation}
\hat X\; = {\mkern 1mu} {G_{\sf CCRNet}}({\mkern 1mu} \left. {Y{\mkern 1mu} } \right|{\mkern 1mu} \hat H{\mkern 1mu} ;\alpha ),
\end{equation}
where ${G_{\sf CCRNet}}( \bullet )$ represents the CCRNet with $\alpha$ being the network parameters. Also, $\hat H$ is the estimated channel matrix, and $Y$is the received signal.

As shown in Fig.\ref{fig:CCRNet}, the CCRNet adopts the paradigm of Generative Adversarial Network (GAN) \cite{goodfellow2014generative} and consists of a generator G and a discriminator D. The generator is trained to recover the transmitted signal and the discriminator D is trained to distinguish whether the recovered signal matches with the input channel condition. Different from \cite{mao2019bilinear} which uses text descriptions as condition input to modulate the given image, here we adopt the estimated channel as the condition input to modulate the received signal and recover the transmitted signal. The proposed CCRNet has two inputs, namely, received signal and estimated channel matrix, which are encoded to feature maps by data encoder and condition encoder, respectively. Then the Bilinear Residual Layer (BRL) is adopted to modulate the received signal to recover the transmitted signal conditioned on the encoded channel feature map. Specifically, three BRL blocks are cascaded with short-cut connections to reconstruct the original data step by step. The architecture of BRL block is shown in the right dashed box of Fig. \ref{fig:CCRNet}, where $V$ represents the encoded condition feature. Also, for BRL1, $U$ represents the encoded data, while for the $i$-th $(i > 1)$  BRL, $U$ represents the output of the $(i-1)$-th BRL. In addition, $\odot $ represents the bilinear operation defined in \cite{mao2019bilinear}, $P$ is the outcome of the bilinear operation, and $ \oplus $ represents the concatenation of two different feature maps along the channel axis. Moreover, short-cut connections are used between the BRL blocks to facilitate the gradient back-propagation and accelerate the convergence of the training process.

\begin{table}[tb]
	\centering
	\caption{CHANNEL AND DATA PARAMETERS}
	\label{tab1}
	\begin{tabular}{ccc}
		\toprule
		\toprule
		Scenario & Vehicular A & Pedestrian A \\
		\midrule
		Carrier Frequency & 2.5GHz & 2.5GHz \\
		Sampling Rate & 1080kHz & 1080kHz \\
		Receiver Speed & 80km/H & 8km/H \\
		Channel Time-Frequency Grid & $72 \times 28$ & $72 \times 28$ \\
		Pilot Arrangement & $18 \times 7$ & $18 \times 7$ \\
		Modulation Scheme & 256-QAM & 256-QAM \\
		A Single OFDM Frame & 72 subcarriers, & 72 subcarriers, \\
		&  28 time slots & 28 time slots \\
		\bottomrule
		\bottomrule
	\end{tabular}
\end{table}

\begin{table}[tb]
    \vspace{-0.15cm}
	\centering
	\caption{NETWORK PARAMETERS}
	\label{tab2}
	\begin{tabular}{p{74 pt}p{60 pt}p{60 pt}}
		\toprule
		\toprule
		Parameter & CENet & CCRNet \\
		\midrule
		Input Size & $18 \times 7 \times 2$ & $72 \times 28 \times 2$ \\
		Output Size & $72 \times 28 \times 2$ & $72 \times 28 \times 2$ \\
		Optimizer & Adam & Adam \\
		Initial Learning rate & 1e-5 & 2e-4 \\
		Batch Size & 16 & 64 \\
		Testing SNR & 10,15,20,25 & 10,15,20 \\
		& 30,35,40 dB & 25,30,35,40 dB \\
		Train Size & $3 \times 30000$ & $3 \times 30000$ \\
		Validation Size & 4000 & 4000 \\
		Test Size & $7 \times 4000$ & $7 \times 4000$ \\
		\bottomrule
		\bottomrule
	\end{tabular}
	\vspace{-0.23cm}
\end{table}

The network architecture of Fusion module is similar with \cite{mao2019bilinear}. Parameters of conv1-4 layers in VGG16\cite{simonyan2014very} are used as the feature extractor for data encoder. The architectures of discriminator, condition encoder and data decoder are shown in TABLE \ref{tab4}.

The training of CCRNet follows from the typical paradigm of GAN \cite{goodfellow2014generative}, which is a min-max game between the generator and the discriminator. Specifically, we train the discriminator
and the generator in an alternating manner similar to the optimization
loss function introduced in \cite{mao2019bilinear}.

\section{EXPERIMENTAL RESULTS}\label{experiment results}
For all simulations, OFDM signals are generated according to the Long Term Evolution (LTE) standard. Also, as in \cite{2012Optimal}, Lattice-type pilot pattern, where pilots are inserted along both time and frequency axes with given period in a diamond-shaped constellation, has been used for pilot transmission. We consider two typical wireless channel models, namely, Vehicular A (VehA) model and Pedestrain A (PedA) model. The corresponding channel parameters are summarized in TABLE \ref{tab1}. Please note, according to [17], for LTE systems operating at the 2.5GHz band, the intercarrier interference (ICI) is negligible when moving at the speed of $80 km/h$. Hence, the impact of ICI is not considered in the current work.

\begin{table}[tp]
	\centering
	\caption{ARCHITECTURE OF DIFFERENT MODULES IN CCRNet}
	\label{tab4}
\begin{tabular}{c|c|c}
	\hline
	\hline
	\multicolumn{3}{c}{Discriminator configuration} \\
	\hline
	layer name & kernel size & stride \\
	\hline
	conv1 & $3 \times 3 \times 4 \times 64$ & 1 \\
	\hline
	BatchNorm1+RELU1 & /     & / \\
	\hline
	conv2 & $3 \times 3 \times 64 \times 128$ & 1 \\
	\hline
	BatchNorm2+RELU2 & / & / \\
	\hline
	conv3 & $3 \times 3 \times 128 \times 1$ & 1 \\
	\hline
	FC    & \multicolumn{2}{c}{Inputsize: $72 \times 28 \times 1$ Outpusize:1}   \\
	\hline
	Sigmoid & / & / \\
	\hline
	\multicolumn{3}{c}{Condition Encoder} \\
	\hline
	layer name & kernel size & stride \\
	\hline
	conv1 & $3 \times 3 \times 2 \times 128$ & 1 \\
	\hline
	BatchNorm1+RELU1 & / & / \\
	\hline
	conv2 & $3 \times 3 \times 128 \times 256$ & 1 \\
	\hline
	BatchNorm2+RELU2 & / & / \\
	\hline
	conv3 & $3 \times 3 \times 256 \times 512$ & 1 \\
	\hline
	BatchNorm3+RELU3 & / & / \\
	\hline
	conv4 & $3 \times 3 \times 512 \times 2$ & 1 \\
	\hline
	\multicolumn{3}{c}{Data Decoder} \\
	\hline
	layer name & kernel size & stride \\
	\hline
	Upsample1 & \multicolumn{2}{c}{$scale factor=2$} \\
	\hline
	Conv1 &  $3 \times 3 \times 512 \times 256$ & 1 \\
	\hline
	BatchNorm1+RELU1 & / & / \\
	\hline
	Upsample2 &  \multicolumn{2}{c}{$scale factor=2$} \\
	\hline
	Conv2 &  $3 \times 3 \times 256 \times 128$ & 1 \\
	\hline
	BatchNorm2+RELU2 & / & / \\
	\hline
	Conv3 &  $3 \times 3 \times 128 \times 2$ & 1 \\
	\hline
	\hline
\end{tabular}
\end{table}

The datasets used for training and testing are generated offline, according to the statstical channel models, namely, VehA model and PedA model. For CENet training, we perform different experiments: one is training the CENet with single SNR value $(22dB\;and\;32dB)$ and the other is training the CENet with a group of SNR values ($\left\{ {10dB,20dB,30dB} \right\}$). As shown in TABLE \ref{tab1}, $90000$ data samples ($30000$ for each SNR) are used for training. $28000$ data samples ($4000$ for each SNR) are used for testing.

The CENet and CCRNet are implemented with Pytorch and python using a GeForce GTX 1080 Ti GPU. All the key parameters are summarized in Table \ref{tab2}.

\begin{figure*}[tbp]
	\centering
	\subfigure[VehA channel, $80km/h$]{
		\begin{minipage}[t]{0.5\textwidth}
			\centering
			\includegraphics[height=5.5cm,width=7cm]{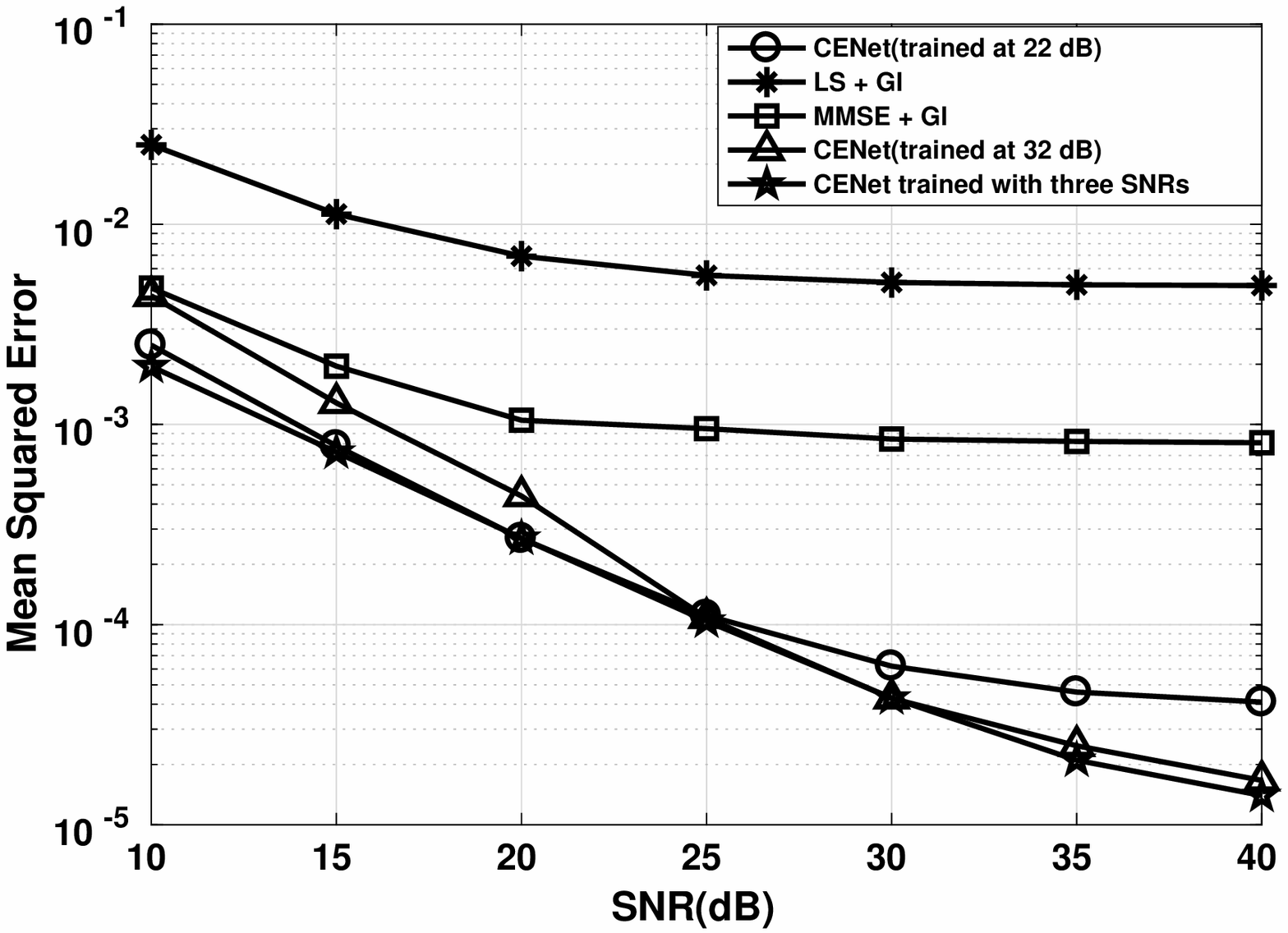}
			\label{VehA channel mse figure}
			\hspace{1cm}
		\end{minipage}%
	}
	\subfigure[PedA channel, $8km/h$]{
		\begin{minipage}[t]{0.5\textwidth}
			\centering
			\includegraphics[height=5.5cm,width=7cm]{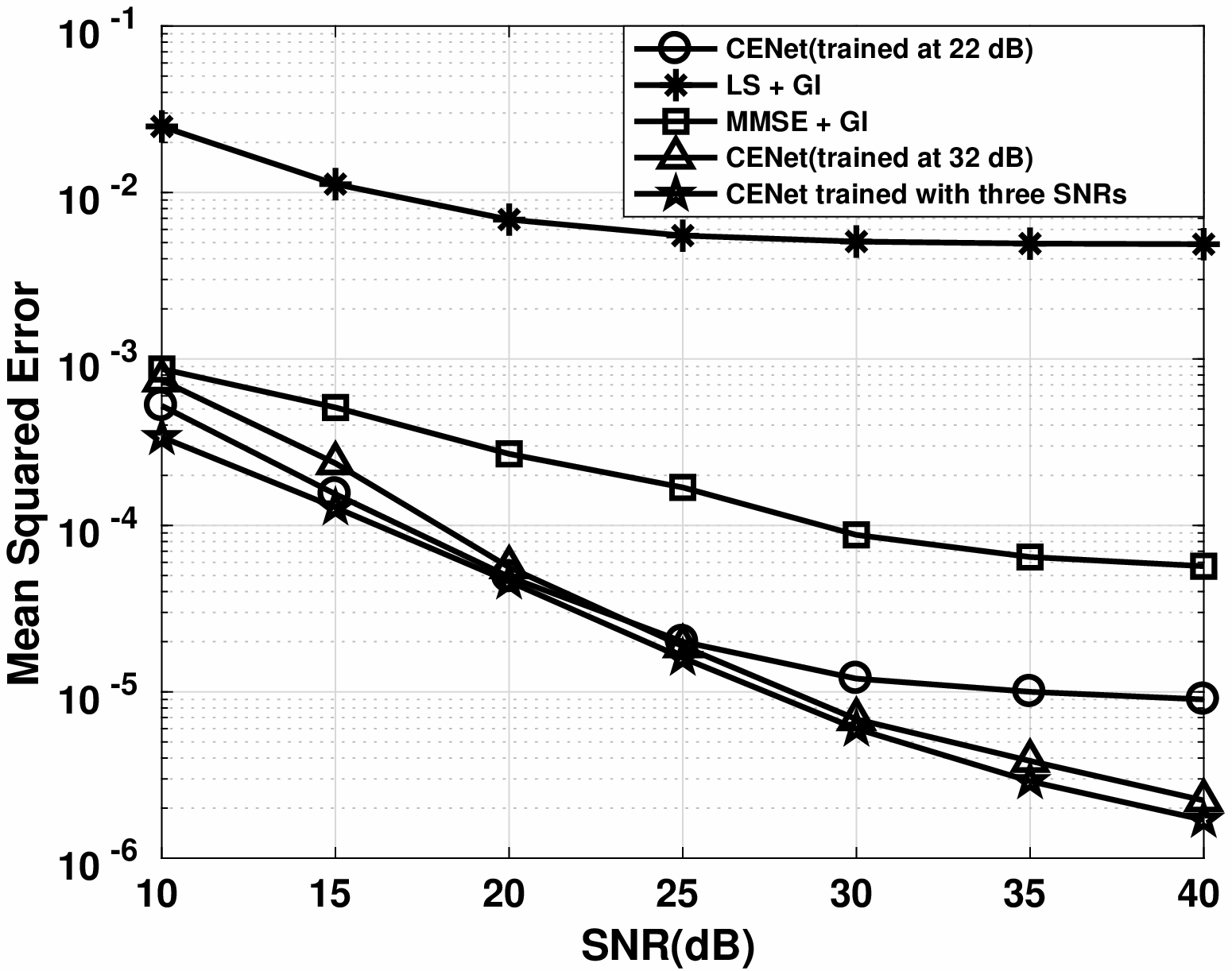}
			\label{PedA channel mse figure}
		\end{minipage}%
	}%
	\caption{The comparison of channel estimation performance for CENet, LS and MMSE schemes}\label{mse of CENet and LS estimator}
\end{figure*}

\begin{figure*}[tp]
	\centering
	\setlength{\belowcaptionskip}{-0.4cm}
	\subfigure[VehA channel, $80km/h$]{
		\begin{minipage}[t]{0.5\textwidth}
			\centering
			\includegraphics[height=5.5cm,width=7cm]{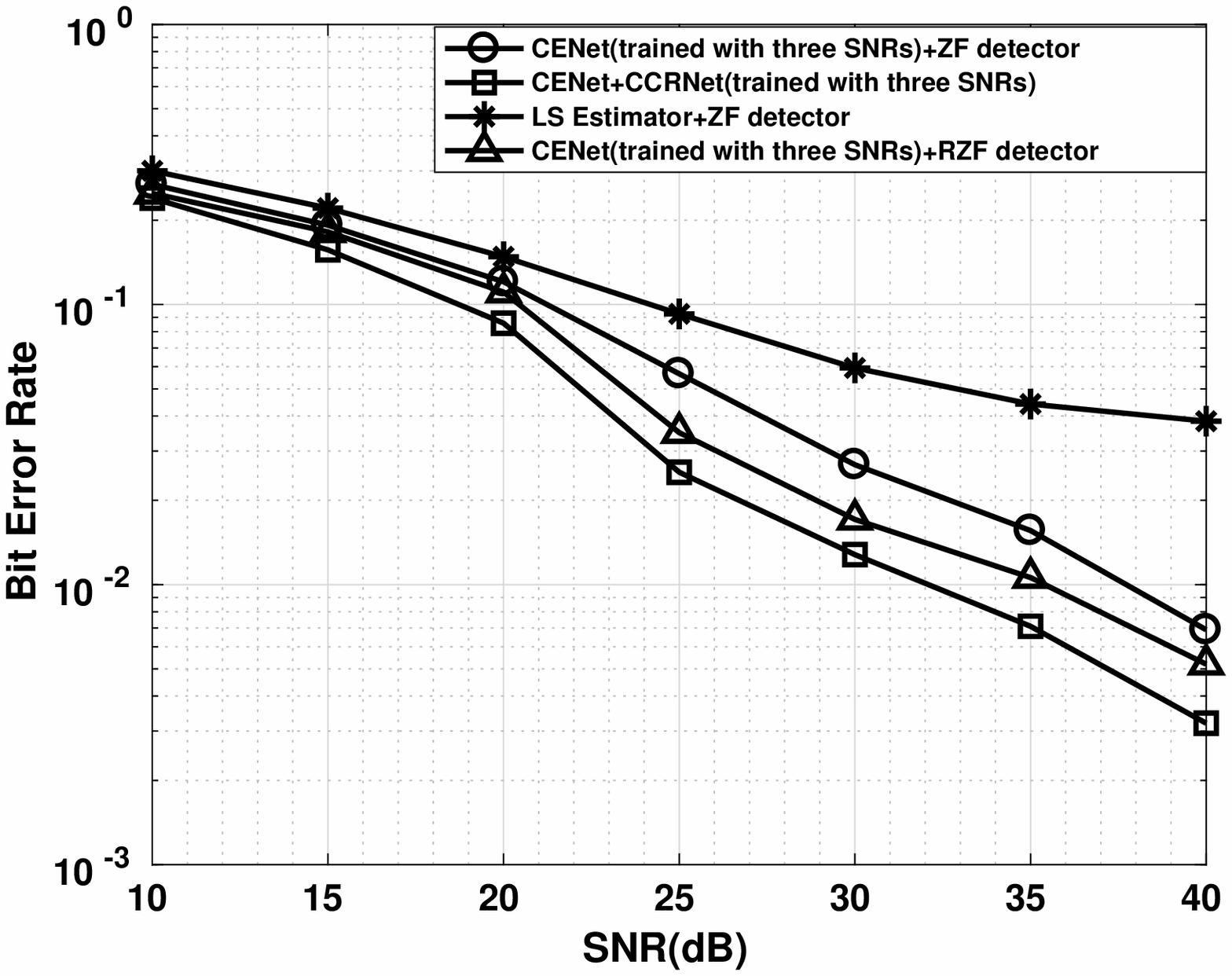}
			\label{VehA channel ber figure}
			\hspace{1cm}
		\end{minipage}%
	}
	\subfigure[PedA channel, $8km/h$]{	
		\begin{minipage}[t]{0.5\textwidth}
			\centering
			\includegraphics[height=5.5cm,width=7cm]{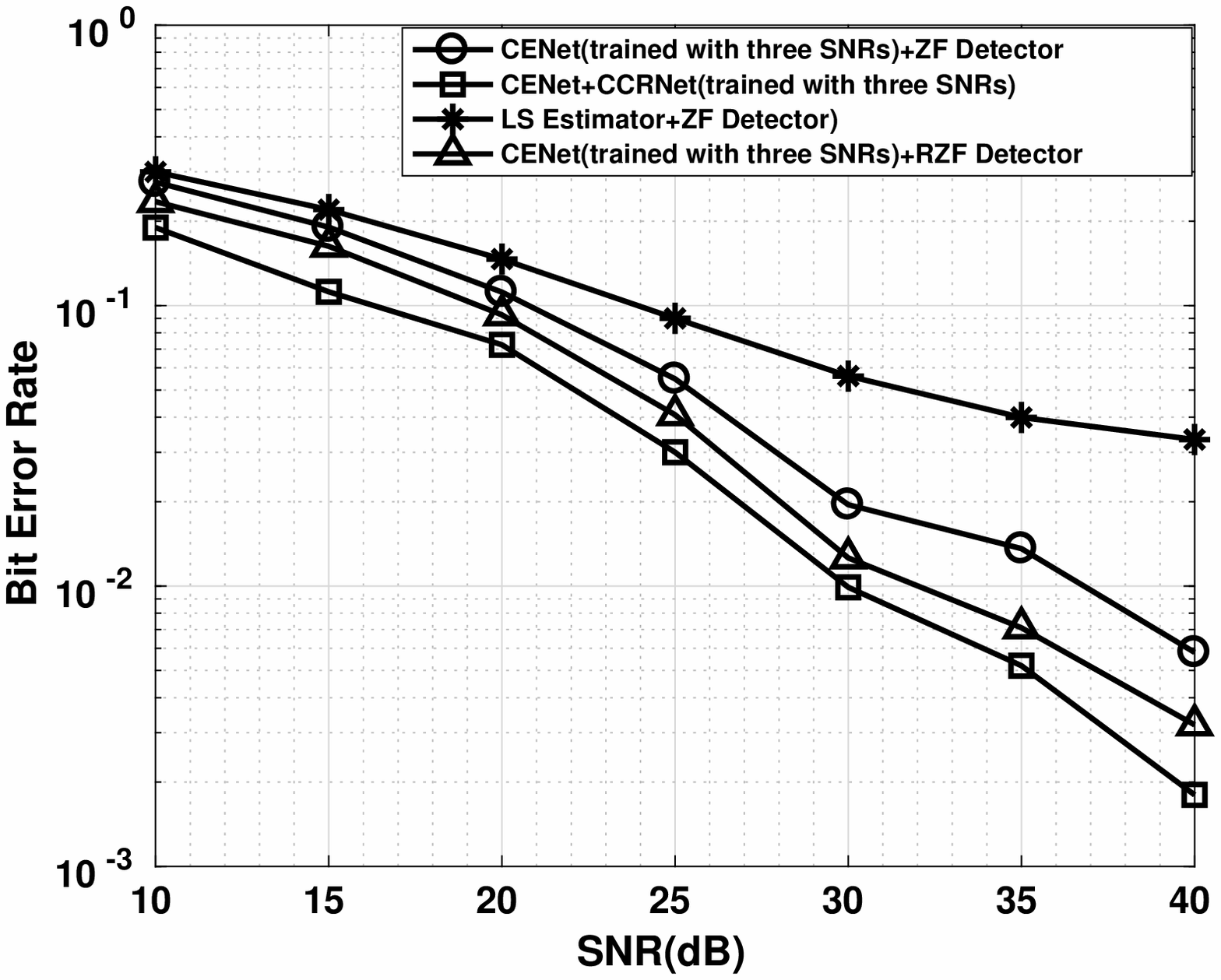}
			\label{PedA channel ber figure}
		\end{minipage}%
	}%
	\caption{The comparison of signal detection performance for CENet with ZF detector, CENet with RZF detector, LS estimator with ZF detector and CENet with CCRNet schemes}\label{ber of CENet an CCRNet}
\end{figure*}

The Mean Square Error (MSE) between the estimated channel output and the corresponding ground-truth channel is adopted to evaluate the performance of different channel estimation schemes. Both the classical LS and MMSE estimation methods adopt Gaussian interpolation (GI) scheme. As illustrated in Fig.\ref{mse of CENet and LS estimator}, the proposed DL-based channel estimation method substantially outperforms the conventional LS channel etimator and MMSE channel estimator in both PedA and VehA scenarios. In addition, we see that the DL-based channel estimation method achieves different performance when trained at different SNRs. For instance, when the operating SNR is smaller than 20 dB, the estimation accuracy is higher if the CENet is trained at 22 dB. In contrast, when the operating SNR is greater than 25 dB, it is better to use the CENet trained at 32 dB. However, it is better to train the model using a group of SNRs. The CENet trained with $\left\{ {10dB,20dB,30dB} \right\}$ achieves lower MSE performance than that trained with single SNR $(22dB\;and\;32dB)$ in all the range of SNR values.


To illustrate the signal detection performance, we compare the proposed CCRNet with the traditional ZF detection scheme and regularized zero forcing (RZF) detection scheme \cite{alkamali2012regularized}. We consider the following four cases: 1) LS for channel estimation and ZF for signal detection; 2) CENet for channel estimation and ZF for signal detection;3) CENet for channel estimation and regularized zero forcing (RZF) for signal detection; 4) CENet for channel estimation and CCRNet for signal recovery. For CCRNet training, a group of SNR values ($\left\{ {10dB,20dB,30dB} \right\}$) is used due to the superior performance shown in the training of CENet. As shown in Fig. \ref{ber of CENet an CCRNet}, the performance curves of different schemes present similar tendency under both VehA and PedA channels. The LS estimator with ZF detector has the highest BER compared with the other schemes. The scheme with CENet and ZF detector achieves lower BER than the scheme with LS estimator and ZF detector. A regularization term is added to avoid the noise enhancement in RZF detection scheme. Thus the scheme with CENet and RZF detector performs better than the scheme with CENet and ZF detecotr, as expected. Nevertheless, The scheme with combined CENet and CCRNet yields the best BER performance, which demonstrates the superior performance of the DL-based approach.

\section{Conclusion}\label{conclusion section}
This paper has proposed a DL-based scheme for joint channel estimation and signal detection in OFDM systems. Exploiting the time and frequency correlation of the underlying wireless channels, a novel CENet is proposed to replace the interpolation procedure by implementing the image super-resolution technique. Subsequently, leveraging on the output of CENet, a CCRNet is designed to recover the transmitted signal. The experimental results demonstrate the effectiveness and superior performance of the proposed DL-based joint channel estimation and signal detection. In addition, the proposed networks are robust to the change of system parameters, thereby making it appealing for practical implementation.


\begin{thebibliography}{99}

\bibitem{hsieh1998channel}
M.-H. Hsieh and C.-H. Wei, ``Channel estimation for ofdm systems
based on comb-type pilot arrangement in frequency selective fading
channels,” {\em IEEE Trans. Consumer Electron.}, vol. 44, no. 1,
pp. 217–225, Feb. 1998.



\bibitem{mahmoud2008kalman}
H. Mahmoud, A. Mousa, and R. Saleem, ``Kalman filter channel estimation based on comb-type pilots for ofdm system in time and frequency selective fading environments,” {\em Proc. of Mosharaka
	International Conference on Communications, Computers and
	Applications}, Amman, 2008, pp: 59–64.

\bibitem{H.Ye0}
H. Ye, G. Y. Li and B.-H. Juang, ``Power of deep learning for channel estimation and signal detection in OFDM systems,'' {\em IEEE Wireless Commun. Letters}, vol. 7, no. 1, pp. 114--117, Feb. 2018.

\bibitem{P.Jiang}
P. Jiang, T. Wang, B. Han, X. Gao, J. Zhang, C-K. Wen, S. Jin and G. Y. Li, ``Artificial intelligence-aided OFDM receiver: Design and experimental results,'' preprint arXiv:1812.06638, 2018.

\bibitem{H.Mao}
H. Mao, H. Lu, Y. Lu and D. Zhu, ``RoemNet: Robust meta learning based channel estimation in OFDM systems,'' in Proc. IEEE International Conference on Communications (ICC), 2019.

\bibitem{J.Zhang}
J. Zhang, C.-K. Wen, S. Jin and G. Y. Li, ``Artificial intelligence-aided receiver for A CP-free OFDM system: Design, simulation, and experimental test,'' preprint arXiv:1903.04766, 2019.

\bibitem{T.Luong}
T. Van Luong, Y. Ko, N. A. Vien, D. H. N. Nguyen and M. Matthaiou, ``Deep learning-based detector for OFDM-IM,'' {\em IEEE Wireless Commun. Letters}, 2019.

\bibitem{Jeff}
E. Balevi and J. Andrews, ``One-Bit OFDM receivers via deep learning,'' {\em IEEE Trans. Commun.}, vol. 67, no. 6, pp. 4326--4336, Jun. 2019.

%
%

%
%


\bibitem{dai2019second}
T. Dai, J. Cai, Y. Zhang, S.-T. Xia, and L. Zhang, ``Second-order
attention network for single image super-resolution,'' {\em Proc. IEEE Conf. Comput. Vis. Pattern Recognit.}, 2019, pp. 11065-11074.

\bibitem{mao2019bilinear}
 X. Mao, Y. Chen, Y. Li, T. Xiong, Y. He, and H. Xue, ``Bilinear
representation for language-based image editing using conditional generative adversarial networks,'' in ICASSP 2019-2019 IEEE International Conference on Acoustics, Speech and Signal Processing (ICASSP), Brighton,2019, pp. 2047–2051.


\bibitem{Chen2004Zero}
S. Chen, G. Dai, and T. Yen, ``Zero-forcing equalization for ofdm
systems over doubly-selective fading channels using frequency domain
redundancy,'' {\em IEEE Trans. Consum. Electron.}, vol. 50,
no. 4, pp. 1004–-1008, Nov.2004.


\bibitem{charrada2017analyzing}
A. Charrada and A. Samet, ``Analyzing performance of joint svr interpolation for lte system with 64-qam modulation under 500 km/h mobile velocity,'' {\em In 2017 Sixth International Conference on Communications and Networking (ComNet)},Béja · Hammamet,2017, pp. 1–6.


\bibitem{goodfellow2014generative}
I. Goodfellow, J. Pouget-Abadie, M. Mirza, B. Xu, D. Warde-Farley, S. Ozair et al., Generative adversarial nets. {\em In Advances in Neural Information Processing Systems}, pp. 2672-2680, 2014.

\bibitem{simonyan2014very}
K. Simonyan, A. Zisserman, ``Very deep convolutional networks for large-scale image recognition,'' {\em Proc. Int. Conf. Learn. Representations}, 2015.

\bibitem{alkamali2012regularized}
F.S. Al-kamali,M.I. Dessouky, B.M. Sallam, F.E. Abd El-Samie, ``Regularized MIMO equalization for SC-FDMA systems,'' {\em CIRC SYST SIGNAL PR.}, vol. 31, no. 4, pp. 1423--1441, Feb. 2012.

\bibitem{2012Optimal}
M. Šimko, Q. Wang, and M. Rupp,``Optimal pilot symbol power allocation under time-variant channels,'' {\em EURASIP J. Wirel. Commun. Netw.}, vol. 2012, no.1, Jul. 2012, doi:10.1186/1687-1499-2012-225.

\bibitem{Rupp2016The}
 M. Rupp, S. Schwarz, and M. Taranetz, The Vienna LTE-Advanced Simulators. Singapore: Springer-Verlag, 2016.
\end{thebibliography}
\end{document}